\documentstyle [10pt]{article}    %{report}
\def\ZzZ{{\hbox{\tenrm Z\kern-.31em{Z}}}}
\def\CcC{{\hbox{\tenrm C\kern-.45em{\vrule height.67em width0.08em depth-
.04em
\hskip.45em }}}}

\newcommand{\lab}{\label}

\newcommand{\bc}{\begin{center}}
\newcommand{\ec}{\end{center}}
\newcommand{\be}{\begin{equation}}
\newcommand{\ee}{\end{equation}}
\newcommand{\bea}{\begin{eqnarray}}
\newcommand{\eea}{\end{eqnarray}}
\newcommand{\bs}{\begin{subequations}}
\newcommand{\es}{\end{subequations}}
\newcommand{\beq}{\begin{eqalignno}}
\newcommand{\eeq}{\end{eqalignno}}

\def\lab{\label}

\def\lab{\label}

\def\IiI{{\hbox{\tenrm I\kern-.19em{I}}}}

\def\uq2{U_q({\uit su}(2))}

\def\CcC{{\hbox{\tenrm C\kern-.45em{\vrule height.67em width0.08em depth-.04em
\hskip.45em }}}}
\def\RrR{{\hbox{\tenrm I\kern-.17em{R}}}}
\def\HhH{{\hbox{\tenrm {I\kern-.18em{H}}\kern-.18em{I}}}}
\def\DdD{{\hbox{\tenrm {I\kern-.18em{D}}\kern-.36em {\vrule height.62em
width0.08em depth-.04em\hskip.36em}}}}

\def\ZzZ{{\hbox{\tenrm Z\kern-.31em{Z}}}}

\def\IiI{{\hbox{\tenrm I\kern-.19em{I}}}}
\def\NnN{{\hbox{\tenrm {I\kern-.18em{N}}\kern-.18em{I}}}}
\def\QqQ{{\hbox{\tenrm {{Q\kern-.54em{\vrule height.61em width0.05em
depth-.04em}\hskip.54em}\kern-.34em{\vrule height.59em width0.05em depth-.04em}}
\hskip.34em}}}
\def\OoO{{\hbox{\tenrm {{O\kern-.54em{\vrule height.61em width0.05em
depth-.04em}\hskip.54em}\kern-.34em{\vrule height.59em width0.05em depth-.04em}}
\hskip.34em}}}

\def\uq2{U_q({\uit su}(2))}

\def\fraz#1#2{{\strut\displaystyle #1\over\displaystyle #2}}

\def\part#1{\fraz{\partial}{\partial#1}}

\def\su2q{SU(2)_q}
\def\h1q{H(1)_q}

\def\nu{N_{1}}

\begin{document}

\bc
{
%\hspace{1cm}
{\bf HOPF ALGEBRA, THERMODYNAMICS AND ENTANGLEMENT\\
IN QUANTUM FIELD THEORY}
\bigskip
\bigskip
%$$ $$
%$$ $$

%\hspace{1cm}

 A. IORIO${}^{1}$, G. LAMBIASE${}^{1}$ AND G. VITIELLO${}^{1,2}$
\bigskip
%\bigskip

%\hspace{1cm}

${}^{1}$Dipartimento di Fisica ``E.R.Caianiello'', Universit\`a di Salerno, 84100
Salerno, Italy\\
and INFN, Gruppo Collegato di Salerno\\
${}^{2}$INFM, Sezione di Salerno\\

iorio@sa.infn.it\\
lambiase@sa.infn.it\\
vitiello@sa.infn.it
$$ $$
}

\ec

%\hspace{1cm}
{\bf Abstract} The quantum deformation of the Hopf algebra describes the skeleton of
quantum field theory, namely its characterizing feature consisting in the existence
of infinitely many unitarily inequivalent representations of the canonical
commutation relations. From this we derive the thermal properties of quantum field
theory, with the entropy playing the role of the generator of the non-unitary time
evolution. The entanglement of the quantum vacuum appears to be robust against
interaction with the environment: on cannot ``unknot the knots" in the infinite
volume limit.

$$  $$

{\bf 1. Introduction}

\bigskip

It is now long known that quantum deformations \cite{25.,20.} of the enveloping
algebras of Lie algebras have a Hopf algebra structure \cite{Celeghini:1991km}. It
has been shown \cite{Iorio:1993jn,Celeghini:1998sy} that dissipative systems, as
well as finite temperature non-equilibrium systems, are properly described in the
frame of the q-deformed Hopf algebra. The analysis of such systems has revealed that
the proper algebraic structure of quantum field theory (QFT) is the deformed Hopf
algebra. The q-parameter acts as a label for the infinitely many unitarily
inequivalent representations (uir) of the canonical commutation relations (ccr). Our
task in this paper is to report about such results on the architecture of QFT and, by
resorting to that, to show its intrinsic thermodynamic nature. We also want to
comment on the robustness of the entanglement of the quantum vacuum against
interaction with the environment \cite{NPB}.

Let us illustrate in a simple way that the Hopf algebra enters in the QFT formalism
since the beginning. We observe that the introduction of the operator algebra
necessary to set up QFT is usually limited to the introduction of the boson
Weyl-Heisenberg (WH) algebra. The additivity of some observables such as the energy,
the momentum and the angular momentum is such an obvious assumption that one does not
even bother to spell it out. It is implicitly given as granted. However, if one is
asked to express it explicitly and formally, then one realizes that the boson WH
algebra is only a part of the full algebraic structure and it becomes natural to
introduce the ``coproduct" map, namely, for the ``addition" of, e.g., the angular
momentum $J^{\alpha}$, ${\alpha} = 1,2,3$, of two particles, $\Delta J^{\alpha} =
J^{\alpha} \otimes {\bf 1} + {\bf 1} \otimes J^{\alpha} \equiv J^{\alpha}_1 +
J^{\alpha}_2$. It is, however, not only in this simple way that one needs the Hopf
structure where the coproduct mapping plays a crucial role. The full algebraic
structure which is needed has to take also into account \cite{Iorio:1994jk} one of
the very special features of QFT, namely the existence of infinitely many
representations of the ccr \cite{BR} (this is a characterizing feature  which  makes
QFT deeply different from quantum mechanics (QM). In QM all the representations of
the ccr are unitary equivalent due to the von Neumann theorem). Then one is led to
consider the q-deformation of the Hopf algebra. Let us see in the following how this
goes.

\bigskip

\noindent {\bf 2. Deformed Hopf algebra and the quantum field theory
structure}

\bigskip

In the following we shall focus on the case of bosons for simplicity. However, our
conclusions can be extended also to fermions \cite{Celeghini:1998sy}.

The coproduct is a homomorphism which duplicates the algebra, ${\Delta}: {\cal A}\to
{\cal A}\otimes {\cal A}$. The operator doubling implied by the coalgebra is a key
ingredient of Hopf algebras. As already said, Lie-Hopf algebras are commonly used in
the familiar addition of energy, momentum and angular momentum. Thus, the operational
meaning of the coproduct is that it provides the prescription for operating on two
modes. Associated to that, there is the {\it doubling} of the degrees of freedom of
the system which is rich of physical meanings (e.g. near a black hole such a doubling
perfectly describes the modes on the two sides of the horizon \cite{NPB,
Martellini:1978sm, Iorio:2001te}).

The bosonic Hopf algebra for a single mode (the case of modes labelled by
the momentum $\kappa$ is straightforward), also called $h(1)$, is generated by the set of
operators $\{ a, a^{\dagger},H,N \}$ with commutation relations:
\be [\, a\, ,\, a^{\dagger} \, ] = \ 2H \, , \quad\; [\, \ N \, ,\, a \, ] = - a \, ,
\quad\; [\, \ N \, ,\, a^{\dagger} \, ] = a^{\dagger} \, , \quad\; [\, \ H \, ,\,
\bullet \, ] = 0 \, ,
\lab{p22}\ee
where $a$ ($a^{\dagger}$) is the generic annihilation (creation) operator and $H$ is
a central operator, constant in each representation. The Casimir operator is given by
${\cal C} = 2NH -a^{\dagger}a$. In ~$h(1)$ the coproduct is defined by
\be \Delta {\cal O} = {\cal O} \otimes {\bf 1} + {\bf 1} \otimes {\cal O} \equiv
{\cal O}_1 + {\cal O}_2 ~,
 \lab{p24}
 \ee
where ${\cal O}$ stands for $a$, $a^{\dagger}$,  $H$ and $N$.  The $q$-deformation of
$h(1)$ is the Hopf algebra $h_{q}(1)$:
\be [\, a_{q}\, ,\,
a_{q}^\dagger \, ] = \ [2H]_{q} \, , \quad\; [\, \ N \, \, , \, a_{q}\, ] = - a_{q}
\, , \quad\;  [ \, \ N \, , \, a_{q}^\dagger \,] = a_{q}^\dagger , \quad\; [\, \ H \,
\, , \,  \bullet \, ] = 0  \, ,
\lab{p26}
\ee
where $N_{q} \equiv N$ and $H_{q} \equiv H$ and $\displaystyle{[x]_{q} = {{q^{x} -
q^{-x}} \over {q - q^{-1}}}}$. The Casimir operator is given by ${\cal C}_{q} =
N[2H]_{q} -a_{q}^{\dagger}a_{q}$.  The coproduct for $a_{q}$ and $a_{q}^\dagger$ now
changes to
\be \Delta a_{q} = a_{q} \otimes {q^{H}} + { q^{-H}} \otimes a_{q} \, , \quad\quad
\Delta a_{q}^{\dagger} = a_{q}^{\dagger} \otimes {q^H} + {q^{-H}} \otimes
a_{q}^{\dagger} ~, ~
\lab{p28}
\ee
while, of course, stays the same for $H$ and $N$.

In the fundamental representation, obtained by setting $H = 1/2$, ${\cal C} = 0$,
$h(1)$ and $h_{q}(1)$ coincide, as it happens for the spin-$\frac{1}{2}$
representation of $su(2)$ and $su_{q}(2)$. The differences appear in the coproduct
and in the higher spin representations. We shall denote by ${\cal F}_{1}$ this
representation space (single mode Fock space).

As customary, one requires $a$ and $a^{\dag}$ ($a_{q}$ and ${a_{q}}^{\dag}$) to be
adjoint operators. This implies that $q$ can only be real or of modulus one. In the
two mode Fock space ${\cal F}_{2} \equiv {\cal F}_{1} \otimes {\cal F}_{1}$, for
$|q|=1$, the hermitian conjugation of the coproduct must be supplemented by the
inversion of the two spaces for consistency with the coproduct homomorphism.
Summarizing,  on ${\cal F}_{2}  = {\cal F}_{1} \otimes {\cal F}_{1}$ it can be
written:
\be \Delta a_{q} =  a_1 q^{1/2} + q^{-1/2} a_2 ~,~~~ ~\Delta
a_{q}^{\dagger} = a_1^{\dagger} q^{1/2}  +q^{-1/2}  a_2^{\dagger} ~, \lab{p213} \ee
\be \Delta H = 1 , ~~~\Delta N =  N_{1} +
 N_{2} ~.  \lab{pdelta} \lab{p214}
\ee

Note that $[a_i , a_j ] = [a_i , a_{j}^{\dagger} ] = 0 , ~ i \neq j $. It is now
possible to show that the full set of infinitely many uir of the ccr in QFT are
classified by use of the deformed Hopf algebra. To do that it is sufficient to show
that the Bogolubov transformations are directly obtained by use of the deformed
copodruct operation. As well known, indeed, the Bogolubov transformations relate
different (i.e. unitary inequivalent) representations. We consider therefore the
following operators (cf. (\ref{p28}) with $H=1/2$ and $q(\theta) \equiv
e^{2\theta}$):
\be {\alpha}_{q(\theta)} \equiv { { {\Delta} a_{q}} \over {\sqrt{[2]_{q}} }} =
{1\over\sqrt{[2]_{q}}} (e^{\theta} a_1 + e^{-\theta} a_2 ) ~,~~~{\beta}_{q(\theta)}
\equiv { 1 \over {\sqrt{[2]_{q}}} } {\delta \over {\delta \theta}} {\Delta} a_{q} =
{1\over\sqrt{[2]_{q}}} (e^{\theta} a_1 -e^{-\theta} a_2 ) \; ,
 \lab{p310} \ee
and h.c.. A set of commuting operators with canonical commutation relations is given
by
\be
{\alpha}(\theta) \equiv
{{\sqrt{[2]_{q}}}  \over 2{\sqrt2}} [ {\alpha}_{q(\theta)} + {\alpha}_{q(- \theta)} -
{\beta}_{q(\theta)}^{\dagger}  +  {\beta}_{q(- \theta)}^{\dagger}] ~, \lab{p314} \ee
\be {\beta}(\theta) \equiv {{\sqrt{[2]_{q}}}  \over 2{\sqrt 2}}[{\beta}_{q(\theta)} +
{\beta}_{q(- \theta)} - {\alpha}_{q(\theta)}^{\dagger}  +  {\alpha}_{q(-
\theta)}^{\dagger} ] ~, \lab{p315} \ee
and h.c. One then introduces
\bea
A(\theta) \equiv \frac{1}{\sqrt{2}} \left ( {\alpha}(\theta ) + {\beta}(\theta
)\right ) &=& A ~{\rm cosh} ~\theta - {B}^{\dagger}
~{\rm sinh} ~\theta ~~, \lab{p321a} \\
B(\theta) \equiv \frac{1}{\sqrt{2}} \left ( {\alpha}(\theta ) - {\beta}(\theta )
\right ) &=& B ~{\rm cosh} ~\theta - A^{\dagger} ~{\rm sinh} ~\theta ~, \lab{p321}
\eea
with $[ A(\theta) , A^{\dagger}(\theta) ] = 1 ~, ~~[ B(\theta) , B^{\dagger}(\theta)
] = 1 $. All other commutators are equal to zero and $A(\theta)$ and $B(\theta)$
commute among themselves. For homogeneity of notation, we have used $A \equiv a_{1}$
and $B \equiv a_{2}$.

Eqs. (\ref{p321a}) and (\ref{p321})  are nothing but the Bogolubov transformations
for the $(A,B)$  pair. In other words, the Bogolubov-transformed operators
$A(\theta)$ and $B(\theta)$ are linear combinations of the coproduct operators
defined in terms of the deformation parameter $q(\theta )$ and of their
${\theta}$-derivatives. Notice that
\be
- i{\delta \over {\delta \theta}} A(\theta) = [{\cal G}, A(\theta)] ~,~~~ -
i{\delta \over {\delta \theta}} B(\theta) = [{\cal G}, B(\theta)] ~,
\lab{p326}\ee
and h.c., where ${\cal G} \equiv -i(A^{\dagger}B^{\dagger} - AB)$ denotes the
generator of (\ref{p321a}) and (\ref{p321}). For a fixed value $\bar{\theta}$, we
have
\be
\exp(i{\bar{\theta}} p_{\theta}) ~A(\theta) = \exp(i{\bar{\theta}}{\cal G})
~A(\theta)~ \exp(-i{\bar{\theta}}{\cal G}) = A( \theta + {\bar {\theta}} ) ~,
\lab{p327}\ee
and similar equations for $B(\theta)$. In eq.(\ref{p327}) the definition
$\displaystyle{p_{\theta} \equiv -i{\delta \over {\delta \theta}}}$ has been used. It
can be regarded as the momentum operator ``conjugate" to the ``degree of freedom"
$\theta$, which thus acquires formal definiteness in the sense of the canonical
formalism.

If $|0\rangle$ denotes the vacuum annihilated by $A$ and $B$, $A|0\rangle = 0 =
B|0\rangle $,  then, at finite volume $V$,
 \begin{equation}\label{19}
 |0 (\theta) \rangle \, = \exp(i \theta {\cal G}) |0 \rangle \, .
  \end{equation}
The vacuum $|0 (\theta) \rangle$ turns out to be an $SU(1,1)$ generalized coherent
state: coherence and the vacuum structure in QFT are thus intrinsically related to
the deformed Hopf algebra.

In the continuum limit in the space of momenta, i.e. in the infinite volume limit,
the number of degrees of freedom becomes uncountable infinite, hence we obtain $
\langle 0(\theta)|0(\theta^{\prime})\rangle\to 0$ ~ as $V\to\infty, \quad \forall
  \theta, \theta^{\prime}, ~ \theta\ne \theta^{\prime}$, thus
the Hilbert spaces ${\cal H_{\theta }}$ and ${\cal H}_{\theta '}$ become unitarily
inequivalent. In this limit, the deformation parameter $\displaystyle{\theta = {1
\over {2}} ~{\rm ln}~q}$ labels the set $\{ {\cal H}_\theta, ~ \forall \theta \}$ of
the infinitely many uir of the ccr
\cite{Iorio:1993jn,Celeghini:1998sy,Celeghini:1993jh}. In conclusion, the deformed
Hopf algebra provides the typical structure one deals with in QFT.

Furthermore, in the case in which the deformation parameter is time-dependent, it
turns out to be related to the so-called heat-term in dissipative systems. This can
be seen by noticing that the Heisenberg equation for $A(t,{\theta} (t))$ is
$$
-i{\dot A}(t,{\theta}(t)) = -i{\delta \over {\delta t}} A(t,{\theta}(t)) -i{{\delta
\theta} \over {\delta t}}~ {\delta  \over {\delta \theta}}A(t, {\theta}(t))=
$$
%\nonumber \\
%
\be \left [ H , A(t,{\theta}(t)) \right ] + {{\delta \theta} \over {\delta t}}~
[{\cal G}, A(t,{\theta}(t)) ] = \left [ H + Q , ~A(t,{\theta}(t)) \right ] ~,
\lab{42} \ee
where $\displaystyle{Q \equiv {{\delta \theta} \over {\delta t}} {\cal G}}$ is the
announced heat-term, and $H$ is the Hamiltonian responsible for the time variation in
the explicit time dependence of $A(t,{\theta}(t))$.  $H + Q$ is therefore to be
identified with the free energy~\cite{Celeghini:1992yv}. Thus, the conclusion is that
variations in time of the deformation parameter actually involve dissipation.

When the proper field description is taken into account, $A$ and $B$ depend on the
momentum $\kappa$ and, as customary in QFT, one deals with the algebras
$\displaystyle{ \bigoplus_{\kappa} h_{\kappa}(1)}$.

\bigskip

\noindent {\bf 3. Entropy, entanglement and environment}
\bigskip

It is remarkable that the ``conjugate momentum" $p_{\theta}$ generates transitions
among the uir (in the infinite volume limit): $\exp (i{\bar \theta}p_{\theta})~
~|0(\theta)> = |0(\theta + {\bar \theta})>$. Use of eq. (\ref{19}) shows that
\be\lab{()} {{\partial}\over{\partial {\theta}_{k}}} |0({\theta})> =  - \left (
{1\over{2}} {{\partial {\cal S_{A}}}\over{\partial {\theta}_{k}}} \right )
|0({\theta})> \quad . \lab{(2.17)} \ee
where, in full generality,  we are using $\theta = {\theta}_{k} $ and
\be\lab{(2.15)}
 S_{A} \equiv - \sum_{\kappa} \Bigl \{ A_{\kappa}^{\dagger} A_{\kappa}
\ln \sinh^{2} \bigl ( {\theta}_{k} \bigr ) - A_{\kappa} A_{\kappa}^{\dagger} \ln
\cosh^{2} \bigl ( {\theta}_{k} \bigr ) \Bigr \} \quad . \ee
Thus $i \left ( {1\over{2}} \hbar {{\partial {\cal S_{A}}}\over{\partial {\theta}}}
\right )$ is the generator of translations in $\theta$ \cite{
Celeghini:1992yv,DeFilippo:1977bk}. The operator $S_{A}$ ($S_{B}$ for the $B$-modes)
is recognized to be the entropy operator
\cite{Celeghini:1992yv,DeFilippo:1977bk,11.}. In dissipative or unstable systems (or
in thermal theories at non-equilibrium) the deformation parameter depends on time,
$\theta = \theta (t)$, and the {\it non-unitary} time evolution is controlled by the
entropy variations \cite{Celeghini:1992yv,DeFilippo:1977bk}. We thus have found that
entropy controls non-unitary time evolution: dissipation implies indeed the choice of
a privileged direction in time evolution ({\it the arrow of time}) with the
consequent breaking of time-reversal invariance. It can be also shown that $S_{A} -
S_{B}$ is a constant in time.

By introducing the { \it free energy} functional for the $A$-modes
\be\lab{(2.14)} {\cal F}_{A} \equiv <0(\theta)| \Bigl ( H_{A} - {1\over{\beta}} S_{A}
\Bigr ) |0(\theta)> \quad , \ee
where $H_{A} = \sum_{\kappa} \hbar \Omega_{\kappa} A_{\kappa}^{\dagger} A_{\kappa}$,
one shows that $ d {\cal F}_{A} = d E_{A} - {1\over{\beta}} d {\cal S}_{A}=0$. This
is the first principle of thermodynamics for a system coupled with environment at
constant temperature and in absence of mechanical work. The change in time  of
particles condensed in the vacuum, $d {\cal N}_{A}$, turns into heat dissipation,
defined as usual ${dQ={1\over{\beta}} dS}$, which recovers the conclusion of the
previous section (cf. eq. (\ref{42}))\cite{Celeghini:1992yv}. These observations thus
lead us to recognize the thermal features of $\{|0(\theta (t))>\}$.

It is also interesting to observe that the thermodynamical arrow of time, whose
direction is defined by the increasing entropy direction, points in the same
direction of the cosmological arrow of time, namely the inflating time direction for
the expanding Universe. This can be shown by considering indeed the quantization of
inflationary models \cite{Alfinito:2000bv}. The concordance between the two arrows of
time is not at all granted and is a subject of an ongoing debate (see, e.g.,
\cite{Hawking:1996ny}).

Let us now turn our attention to the entanglement. It is convenient to consider the
case in which we have two modes, say $A$ and $\bar A$ (as, for example, in the case
of a complex scalar field). The corresponding {\it doubled} modes are $B$ and $\bar
B$. For typographical simplicity, we denote them as $A \equiv A^{(+)}$, ${\bar A}
\equiv {\bar A}^{(+)}$, $B \equiv A^{(-)}$ and ${\bar B} \equiv {\bar A}^{(-)}$. We
also use $\sigma = \pm$. Then the vacuum can be expressed as a $SU(1,1) \times
SU(1,1)$ generalized coherent state \cite{PER} of Cooper-like pairs
 \begin{equation}\label{26}
 |0({\theta}) \rangle = \,G(\theta)|0 \rangle = \,
\frac{1}{Z}\,\exp\left[{\sum_{\sigma} \sum_{k} \;\tanh\theta_{k}
 A_k^{(\sigma)\dagger} \bar{A}_{k}^{(-\sigma)\dagger}}\right]\, |0\rangle\,{,}
 \end{equation}
where $Z= \prod_k\;\cosh^2\theta_{k}$. The state
$|0 (\theta) \rangle$ in (\ref{26}), can be
rewritten as
\begin{equation}
  |0(\theta) \rangle = \frac{1}{Z} \left[ |0 \rangle  + \sum_k \;\tanh\theta_{k}
  \left( | 1^{(+)}_k , \bar{0} \rangle \otimes |0, \bar{1}^{(-)}_{k} \rangle
  + | 0, \bar{1}^{(+)}_{k} \rangle \otimes |1^{(-)}_{ k} , \bar{0} \rangle
  \right)  + \dots \right] \,, \label{expans-vacuum}
\end{equation}
where, we denote by $|n^{(\sigma)}_k, \bar{m}^{(\sigma)}_{k}\rangle$ a state of $n$
particles and $m$ ``antiparticles" in whichever sector $(\sigma)$. Note that for the
generic $n^{\rm th}$ term, the state $| n^{(\sigma)}_k , \bar{0} \rangle \equiv
|1^{(\sigma)}_{k_1}, \ldots, 1^{(\sigma)}_{k_n}, \bar{0} \rangle$, and similarly for
antiparticles.

By introducing a well known notation, $\uparrow$ for a particle, and $\downarrow$ for
an antiparticle, the two-particle state in (\ref{expans-vacuum}) can be written as
\begin{equation}\label{enta-onestate}
| \uparrow^{(+)} \rangle \otimes | \downarrow^{(-)} \rangle + | \downarrow^{(+)}
\rangle \otimes | \uparrow^{(-)} \rangle \,,
\end{equation}
which is an entangled state of particle and antiparticle living in the two sectors
$(\pm)$. The generic $n^{\rm th}$ term in (\ref{expans-vacuum}) shares exactly the
same property as the two-particle state, but this time the $\uparrow$ describes a
{\it set} of $n$ particles, and $\downarrow$ a {\it set} of $n$ antiparticles. The
mechanism of the entanglement, induced by the q-deformation, takes place at all
orders in the expansion, always by grouping particles and antiparticles into two
sets. Thus the whole vacuum $|0(\theta) \rangle$ is an infinite superposition of
entangled states\footnote{A similar structure also arises in the
temperature-dependent vacuum of Thermo-Field Dynamics \cite{11.} (see also
\cite{Song}).}
\begin{equation}\label{enta-series}
  |0(\theta) \rangle = \sum_{n=0}^{+\infty} \sqrt{W_n} |{\rm Entangled} \rangle_n
  \,,
\end{equation}
\begin{equation}\label{wn}
  W_n = \prod_k
  \frac{\sinh^{2n_k}\theta_{k}}{\cosh^{2(n_k+1)}\theta_{k}}\,,
\end{equation}
with $ 0< W_n <1 \quad {\rm and} \quad  \sum_{n=0}^{+\infty} W_n = 1$. The
coefficients $\sqrt{W_n}$ of the expansion in Eq. (\ref{enta-series}) appear also in
the discussion about the entropy of the black hole \cite{NPB, Iorio:2001te}.

Of course, the probability of having entanglement of two sets of $n$ particles and
$n$ antiparticles is $W_n$. At finite volume, being $W_n$ a decreasing monotonic
function of $n$, the entanglement is suppressed for large $n$. It appears then that
only a finite number of entangled terms in the expansion (\ref{enta-series}) is
relevant. Nonetheless this is only true at finite volume (the QM limit), while the
interesting case occurs in the infinite volume limit, which one has to perform in a
QFT setting.

The entanglement is generated by $G(\theta)$, where the field modes in one sector
$(\sigma)$ are coupled to the modes in the other sector $(-\sigma)$ via the
deformation parameter $\theta_{k}$. Since the deformation parameter may describe the
background gravitational field (environment) \cite{Iorio:2001te}, the temperature
effects \cite{Celeghini:1998sy}, the dissipative effects \cite{Iorio:1993jn} and
other effects of the environment in which the system is embedded, it appears that the
origin of the entanglement {\it is} the environment, in contrast with the usual QM
view, which attributes to the environment the loss of the entanglement. In the
present treatment such an origin for the entanglement makes it quite robust.

One further reason for the robustness is that this entanglement is realized in the
limit to the infinite volume {\it once and for all} since then there is no unitary
evolution to disentangle the vacuum: at infinite volume one cannot "unknot the
knots". Such a non-unitarity is only realized when {\it all} the terms in the series
(\ref{enta-series}) are summed up, which indeed happens in the $V\to \infty$ limit.

As a final comment let us observe that the doubling implied by the deformed Hopf
algebra has been useful in several applications, ranging from unstable particles
\cite{DeFilippo:1977bk}, to coherence in quantum Brownian motion
\cite{Blasone:1998xt}, squeezed states in quantum optics \cite{Celeghini:1993jh},
topologically massive theories in the infrared region in 2+1 dimensions
\cite{Blasone:1996yh}, the Chern-Simons-like dynamics of Bloch electrons in solids
\cite{Blasone:1996yh}, the quantization of matter in curved background
\cite{NPB,Martellini:1978sm, Iorio:2001te}. These features are also common to
two-dimensional gravity models \cite{Cangemi:1996yz} and to the study of quantization
arising from the loss of information \cite{'tHooft:1999gk}. Moreover, it has been
applied \cite{Vitiello:1995wv} to the study of the memory capacity problem in the
quantum model for the brain.

%\newpage

%%%%%%%%%%%%%%%%%%%%%%%%%%%%%%%%%%%%%%


\begin{thebibliography}{99}

\bibitem{25.} Drinfeld V.G., in {\it Proc. ICM Berkeley, CA},
A.M. Gleason, ed,; AMS,
       Providence, R.I., 1986, 798p \\
M.Jimbo,   Int. J. of Mod. Phys. {
       \bf A4}, 3759 (1989).  \\
Yu.I.Manin, {\it Quantum groups and Non-Commutative Geometry},
        CRM, Montreal, 1988.


\bibitem{20.}   L.C.Biedenharn, J.Phys.  {\bf A22}, L873 (1989).\\
A.J.Macfarlane, J. Phys.  {\bf A22}, 4581
        (1989).


%\bibitem{26.}
%\cite{Celeghini:1991km}
\bibitem{Celeghini:1991km}
E.~Celeghini, T.~D.~Palev and M.~Tarlini,
%``The quantum superalgebra B-q(0/1) and q deformed
%creation and annihilation operators,''
Mod.\ Phys.\ Lett.\ B {\bf 5}, 187 (1991).\\
%%CITATION = MPLAE,B5,187;%%
%\cite{Kulish:1989sv}
%\bibitem{Kulish:1989sv}
P.~P.~Kulish and N.~Y.~Reshetikhin,
%``Universal R Matrix Of The Quantum Superalgebra Osp(2 $|$ 1),''
Lett.\ Math.\ Phys.\  {\bf 18}, 143 (1989).
%%CITATION = LMPHD,18,143;%%



%\bibitem{5.}
%\cite{Iorio:1993jn}
\bibitem{Iorio:1993jn}
%A.~Iorio and G.~Vitiello,
%``Quantization of damped harmonic oscillator, thermal
%field theory and q groups,''
%in {\it Proceedings of The Third
 %       International Workshop on Thermal Field Theories - Banff/CAP
  %      Workshop on Thermal field theory}, Eds. F.C.Khanna et al.,
   %     World Scientific, Singapore 1994, p.71;
%arXiv:math-ph/0009036.\\
%%CITATION = MATH-PH 0009036;%%
%\cite{Iorio:1995fy}
%\bibitem{Iorio:1995fy}
A.~Iorio and G.~Vitiello,
%``Quantum dissipation and quantum groups,''
Annals Phys.\  {\bf 241}, 496 (1995)
%[arXiv:hep-th/9503136].
%%CITATION = HEP-TH 9503136;%%



%\bibitem{41.}
%\cite{Celeghini:1998sy}
\bibitem{Celeghini:1998sy}
E.~Celeghini, S.~De Martino, S.~De Siena, A.~Iorio, M.~Rasetti and G.~Vitiello,
%``Thermo field dynamics and quantum algebras,''
Phys.\ Lett.\ A {\bf 244}, 455 (1998).
%[arXiv:hep-th/9801031].
%%CITATION = HEP-TH 9801031;%%
%\cite{DeMartino:1996zm}
%\bibitem{DeMartino:1996zm}
%S.~De Martino, S.~De Siena and G.~Vitiello,
%``Quantum groups and thermo field dynamics,''
%Int.\ J.\ Mod.\ Phys.\ B {\bf 10}, 1615 (1996).
%%CITATION = IMPAE,B10,1615;%%

\bibitem{NPB}
A.~Iorio, G. ~Lambiase and G.~Vitiello, ``Black hole entropy, entanglement and
holography", arXiv:hep-th/0204034


%\bibitem{21.}
%\cite{Iorio:1994jk}
\bibitem{Iorio:1994jk}
A.~Iorio and G.~Vitiello,
%``Quantum groups and von Neumann theorem,''
Mod.\ Phys.\ Lett.\ B {\bf 8}, 269 (1994)
%[arXiv:math-ph/0009035].
%%CITATION = MATH-PH 0009035;%%



%cite{10.}
\bibitem{BR} O.Bratteli and D.W.Robinson, {\it
    Operator Algebras and Quantum
Statistical Mechanics}, Springer, Berlin, 1979.


%\bibitem{28.}
%\cite{Martellini:1978sm}
\bibitem{Martellini:1978sm}
M.~Martellini, P.~Sodano and G.~Vitiello,
%``Vacuum Structure For A Quantum Field Theory In Curved Space-Time,''
Nuovo Cim.\ A {\bf 48}, 341 (1978).
%%CITATION = NUCIA,A48,341;%%


%\cite{Iorio:2001te}
\bibitem{Iorio:2001te}
A.~Iorio, G.~Lambiase and G.~Vitiello,
%``Quantization of scalar fields in
%curved background and quantum  algebras,''
Annals of Phys. {\bf 294}, 234 (2001).
%arXiv:hep-th/0104162.
%%CITATION = HEP-TH 0104162;%%


%cite{27.}
%\bibitem{27.}
%\cite{Celeghini:1993jh}
\bibitem{Celeghini:1993jh}
%E.~Celeghini, S.~De Martino, S.~De Siena, M.~Rasetti and G.~Vitiello,
%``Quantum groups, squeezing, Bloch and theta functions,''
%Mod.\ Phys.\ Lett.\ B {\bf 7}, 1321 (1993).\\
%%CITATION = MPLAE,B7,1321;%%
%\cite{Celeghini:1995jh}
%\bibitem{Celeghini:1995jh}
E.~Celeghini, S.~De Martino, S.~De Siena, M.~Rasetti and G.~Vitiello,
%``Quantum groups, coherent states, squeezing and
%lattice quantum mechanics,''
Annals Phys.\  {\bf 241}, 50 (1995).\\
%[arXiv:hep-th/9310132].
%%CITATION = HEP-TH 9310132;%%
%cite{29.}
%\bibitem{29.}
%\cite{Celeghini:1991jw}
%\bibitem{Celeghini:1991jw}
E.~Celeghini, M.~Rasetti and G.~Vitiello,
%``On squeezing and quantum groups,''
Phys.\ Rev.\ Lett.\  {\bf 66}, 2056 (1991).
%%CITATION = PRLTA,66,2056;%%



%\bibitem{3.}
%\cite{Celeghini:1992yv}
\bibitem{Celeghini:1992yv}
E.~Celeghini, M.~Rasetti and G.~Vitiello,
%``Quantum dissipation,''
Annals Phys.\  {\bf 215}, 156 (1992).
%%CITATION = APNYA,215,156;%%


%\bibitem{24.}
%\cite{DeFilippo:1977bk}
\bibitem{DeFilippo:1977bk}
S.~De Filippo and G.~Vitiello,
%``Vacuum Structure For Unstable Particles,''
Lett.\ Nuovo Cim.\  {\bf 19}, 92 (1977).
%%CITATION = NCLTA,19,92;%%

%cite{11.}
\bibitem{11.}
Y. Takahashi and H. Umezawa,  Collective Phenomena {\bf 2}, 55 (1975). \\
H.Umezawa, {\it Advanced field theory: micro, macro and thermal concepts}, AIP, N.Y.
1993.

%\bibitem{42.}
%\cite{Alfinito:2000bv}
\bibitem{Alfinito:2000bv}
E.~Alfinito, R.~Manka and G.~Vitiello,
%``Vacuum structure for expanding geometry,''
Class.\ Quant.\ Grav.\  {\bf 17}, 93 (2000).
%[arXiv:gr-qc/9904027].
%%CITATION = GR-QC 9904027;%%
%\cite{Alfinito:1999gg}
%\bibitem{Alfinito:1999gg}
%E.~Alfinito and G.~Vitiello,
%``Canonical quantization and expanding metrics,''
%Phys.\ Lett.\ A {\bf 252}, 5 (1999).
%[arXiv:gr-qc/9807002].
%%CITATION = GR-QC 9807002;%%




%\bibitem{43.}
%\cite{Hawking:1996ny}
\bibitem{Hawking:1996ny}
S.~W.~Hawking and R.~Penrose,
%``The Nature of space and time,''
Sci.\ Am.\  {\bf 275}, 44 (1996).
%%CITATION = SCAMA,275,44;%%


%cite{23.}
\bibitem{PER} A.Perelomov, {\it Generalized Coherent States and Their
        Applications}, Springer,
         Berlin, 1986.

\bibitem{Song} D.Mi, H.S.Song, and Y.An,
Mod. Phys. Lett. {\bf A 16}, 655 (2001).


%cite{38.}
%\bibitem{38.}
%\cite{Blasone:1998xt}
\bibitem{Blasone:1998xt}
M.~Blasone, Y.~N.~Srivastava, G.~Vitiello and A.~Widom,
%``Phase coherence in quantum Brownian motion,''
Annals Phys.\  {\bf 267}, 61 (1998).\\
%[arXiv:quant-ph/9707048].
%%CITATION = QUANT-PH 9707048;%%
%\bibitem{4.}
%\cite{Srivastava:1995yf}
%\bibitem{Srivastava:1995yf}
Y.~N.~Srivastava, G.~Vitiello and A.~Widom,
%``Quantum dissipation and quantum noise,''
Annals Phys.\  {\bf 238}, 200 (1995)
%[arXiv:hep-th/9502044].
%%CITATION = HEP-TH 9502044;%%


%cite{39.}
%\bibitem{39.}
%\cite{Blasone:1996yh}
\bibitem{Blasone:1996yh}
M.~Blasone, E.~Graziano, O.~K.~Pashaev and G.~Vitiello,
%``Dissipation and topologically massive gauge theories
%in the  pseudo-Euclidean plane,''
Annals Phys.\  {\bf 252}, 115 (1996).
%[arXiv:hep-th/9603092].
%%CITATION = HEP-TH 9603092;%%




%cite{40.}
%\bibitem{40.}
%\cite{Cangemi:1996yz}
\bibitem{Cangemi:1996yz}
D.~Cangemi, R.~Jackiw and B.~Zwiebach,
%``Physical states in matter coupled dilaton gravity,''
Annals Phys.\  {\bf 245}, 408 (1996).
%[arXiv:hep-th/9505161].
%%CITATION = HEP-TH 9505161;%%


%\bibitem{32.}
%\cite{'tHooft:1999gk}
\bibitem{'tHooft:1999gk}
G.~'t Hooft,
%``Quantum gravity as a dissipative deterministic system,''
Class.\ Quant.\ Grav.\  {\bf 16}, 3263 (1999).\\
%[arXiv:gr-qc/9903084].
%%CITATION = GR-QC 9903084;%%
%\cite{'tHooft:2001ct}
%\bibitem{'tHooft:2001ct}
G.~'t Hooft, ``Quantum mechanics and determinism,''
arXiv:hep-th/0105105.\\
%%CITATION = HEP-TH 0105105;%%
%\cite{'tHooft:2001ar}
%\bibitem{'tHooft:2001ar}
G.~'t Hooft, ``Determinism in free bosons,'' arXiv:hep-th/0104080.\\
%%CITATION = HEP-TH 0104080;%%
%\bibitem{33.}
%\cite{Blasone:2001ew}
%\bibitem{Blasone:2001ew}
M.~Blasone, P.~Jizba and G.~Vitiello,
%``Dissipation and quantization,''
Phys.\ Lett.\ A {\bf 287}, 205 (2001)
%[arXiv:hep-th/0007138].
%%CITATION = HEP-TH 0007138;%%



%\bibitem{35.}
%\cite{Vitiello:1995wv}
\bibitem{Vitiello:1995wv}
G.~Vitiello,
%``Dissipation and memory capacity in the quantum brain model,''
Int.\ J.\ Mod.\ Phys.\ B {\bf 9}, 973 (1995).\\
%[arXiv:quant-ph/9502006].
%%CITATION = QUANT-PH 9502006;%%
%\bibitem{36.}
%\cite{Alfinito:2000ck}
%\bibitem{Alfinito:2000ck}
E.~Alfinito and G.~Vitiello,
%``Formation and life-time of memory domains in the
%dissipative quantum model of brain,''
Int.\ J.\ Mod.\ Phys.\ B {\bf 14}, 853 (2000) [Erratum-ibid.\ B {\bf 14}, 1613
(2000)]\\
%[arXiv:quant-ph/0002014].
%%CITATION = QUANT-PH 0002014;%%
%\bibitem{37.}
G.Vitiello, {\it My Double unveiled - The dissipative quantum model of brain}, John
Benjamins Publ. Co., Philadelphia, Amsterdam 2001.


\end{thebibliography}
\end{document}